\documentstyle[10pt,psfig]{article}
\begin{document}

\title{X-ray imaging and spectroscopy of (radio-quiet) AGN: 
        highlights from {\itshape{Chandra}} and {\itshape XMM-Newton} }
\author{Stefanie Komossa \\
 Max-Planck-Institut f\"ur extraterrestrische Physik, Giessenbachstr., \\
D-85748 Garching, Germany; skomossa@mpe.mpg.de
}


\date{}
\maketitle

\vspace*{-6.5cm}
{\small{\bf{Review, to appear in {\itshape{China-Germany Workshop on 
the Multiwavelength View on AGN}}
 (Lijiang, Aug. 2002); Yunnan Observatory Reports,
  J. Wei, F. Cheng, G. Hasinger et al. (eds), in press}}}

\vspace*{5.0cm}

\begin{abstract}
The X-ray observatories {\sl XMM-Newton} and {\sl Chandra}
provided a wealth of exciting new results. {\sl Chandra} delivered
X-ray images of outstanding detail, reaching  subarcsecond
spatial resolution for the first time in X-ray astronomy.
{\sl XMM-Newton} provided the highest-signal X-ray
spectra ever taken.
The imaging and spectral observations are greatly improving our
understanding of the physical processes in the central region
of active galaxies. The increased sensitivity also allows
to peer into the high-redshift universe beyond $z$=5.

I shall provide a review of recent X-ray highlights on (radio-quiet)
AGN from low to high redshift, and exciting questions still open.
I conclude with a glimpse into the future of X-ray astronomy.

\end{abstract}

\section{Introduction}

\subsection{X-ray probes of the black hole region of AGN}

X-ray emission
originates from the immediate vicinity of
the black hole.
The detection of luminous, hard, power-law-like X-ray emission,
rapid variability, and the recent discovery of 
relativistic effects in the iron-K line profile 
provided excellent evidence for the presence of supermassive black holes (SMBHs)  
in active galaxies.
X-ray observations currently constitute the most powerful
means to explore the black hole region of
AGN.

X-rays at the centers of AGN arise in the
accretion-disk -- corona system.
On larger scales, but still within the central region, X-rays might be emitted
by a hot intercloud medium at distances of the broad or narrow-line region. 
The X-rays which originate from the accretion-disk region
are reprocessed in form of absorption and partial re-emission
(e.g., George \& Fabian 1991, Netzer 1993, Krolik \& Kriss 1995,
Collin-Souffrin et al. 1996, Komossa \& Fink 1997)
as they
make their way out of the nucleus.
The reprocessing bears the disadvantage of veiling the {\em intrinsic}
X-ray spectral shape,
and the spectral disentanglement of many different potentially contributing
components is not always easy.
However, reprocessing also offers the unique chance
to study the physical conditions
and dynamical states of the reprocessing material, including 
the outer parts of the accretion disk; the ionized absorber;
the torus, which plays an important role in AGN unification schemes
(Antonucci 1993); and the BLR and NLR.
Detailed modeling of the reprocessor(s) is also indispensable  
to recover the shape and properties of the {\em intrinsic} X-ray spectrum.

\subsection{{\itshape{Chandra}} and {\itshape{XMM-Newton}}  }

There are two key abilities 
of the new X-ray observatories {\sl XMM-Newton} (e.g., Jansen et al. 2001)
and {\sl Chandra} (e.g., Tananbaum \& Weisskopf 2001, Weisskopf et al. 2002).
One is the superb spatial resolution of {\sl Chandra}, 
providing images rich in details on the 1\,arcsec-level
for the first time. 
The other is the increase in spectral resolution of 
{\sl XMM} and {\sl Chandra}, providing X-ray spectra
with such a wealth of features including many narrow absorption
and emission lines, detected in X-rays for the first time.      

Both missions have imaging detectors (energy bandpass $\sim$(0.1--10) keV)
and
grating spectrometers aboard. 
Below, a short review of results 
from these observatories is given.
Due to space limitations it will necessarily be incomplete, both
in topics covered and in citations given. My apologies in advance.

\section{X-ray imaging spectroscopy of AGN}

\subsection{Nearby Seyfert galaxies}

Among the first published AGN X-ray images taken with {\sl Chandra}
was that of NGC\,1068 (Young et al. 2001).  
The image demonstrated the outstanding spatial
resolution power of the X-ray observatory.
While it had been noted before that the X-ray emission
of NGC\,1068 was extended,
with {\sl Chandra} for the first time a wealth of details was seen. One could
actually immediately realize one was looking at a {\em galaxy} just
by inspecting the X-ray image (Fig. 1). 
The following details could be recognized (Young et al. 2001):
a bright compact core (165\,pc),
emission extending to the north-east coincident with the NLR,
and large-scale emission partly tracing the spiral arms.
Similarly spectacular is the X-ray spectrum of NGC\,1068,
addressed below (Sect. 3.1).

The nearby Seyfert galaxy NGC\,4151 is 
among the best studied AGN across the
electromagnetic
spectrum. Again, its extended X-ray emission 
was noted previously (e.g., Elvis et al. 1983). 
Details remained unknown, though. 
{\sl Chandra} resolved the extended emission (Ogle et al. 2000),
showed that it is spatially coincident with the NLR and
of high temperature.

\subsection{Ultra-luminous infrared galaxies}
Ultra-luminous infrared galaxies (ULIRGs)
are characterized by their huge luminosity output
in the infrared, powered by central starburst and/or AGN activity. 
X-rays provide an excellent tool to search for
deeply obscured AGN at the centers of ULIRGs, and also 
allow to trace starburst-superwind activity.  
With {\sl Chandra}, 
the ULIRGs Mrk\,273 (Xia et al. 2002), Mrk 231 (Gallagher et al. 2002b)
and Arp 220 (Clements et al. 2002)
were resolved in X-rays, revealing many details.  
Since there is a separate review on ULIRGs
(Xia, these proceedings), 
the rest of this section will concentrate on
NGC\,6240, a recent {\sl Chandra} observation of which revealed
some surprises: Despite decades of searches and debates
whether an AGN is present at all in this source (for an overview
see Sect. 2 of Komossa \& Schulz 1999), a clear detection remained elusive
prior to the registration of hard X-ray emission beyond 2-10 keV  
with {\sl ROSAT}, {\sl ASCA}, and particularly {\sl BeppoSAX} and {\sl RXTE}.  
Both, {\sl Beppo-Sax} and {\sl RXTE} did not 
posses spatial resolution across the field of view,
but the most likely counterpart of the hard X-ray emission
was argued to be NGC 6240. The unprecedented
spatial resolution of {\sl Chandra}, coupled with simultaneous energy
information, let to the detection of 
luminous, hard X-ray emission and strong neutral iron lines 
from {\em both} {\em cores} of the galaxy (Fig. 1,2). These properties are the characteristic
features of the presence of AGN, implying {\em both} cores of NGC\,6240
are active (Komossa et al. 2003).
The two supermassive black holes are presently separated by $\sim$3000\,ly.
The final merging of these black holes is expected to produce 
a strong gravitational wave signal of the kind detectable with 
the future space-borne observatory LISA (e.g., Danzmann 1996, Centrella 2003).     

In addition to the double black hole, NGC\,6240 exhibits luminous extended
X-ray emission, first detected with the instruments aboard {\sl ROSAT}
(e.g., Schulz et al. 1998), and spatially and spectroscopically
resolved with {\sl Chandra} (Komossa et al. 2003). It is very likely 
related to starburst-superwind activity and changes its
rich structure in dependence of energy. While the more extended starburst
loops have a lower X-ray temperature, there is also a more compact 
component surrounding the two nuclei which is of higher temperature
and shows a tendency for higher metal abundances than the more
extended component.  

{\sl Chandra} observations of more ULIRGs are expected to provide 
a sharp and detailed look at the very centers of these galaxies
in the coming years. 

\begin{figure*}[bt]
\psfig{file=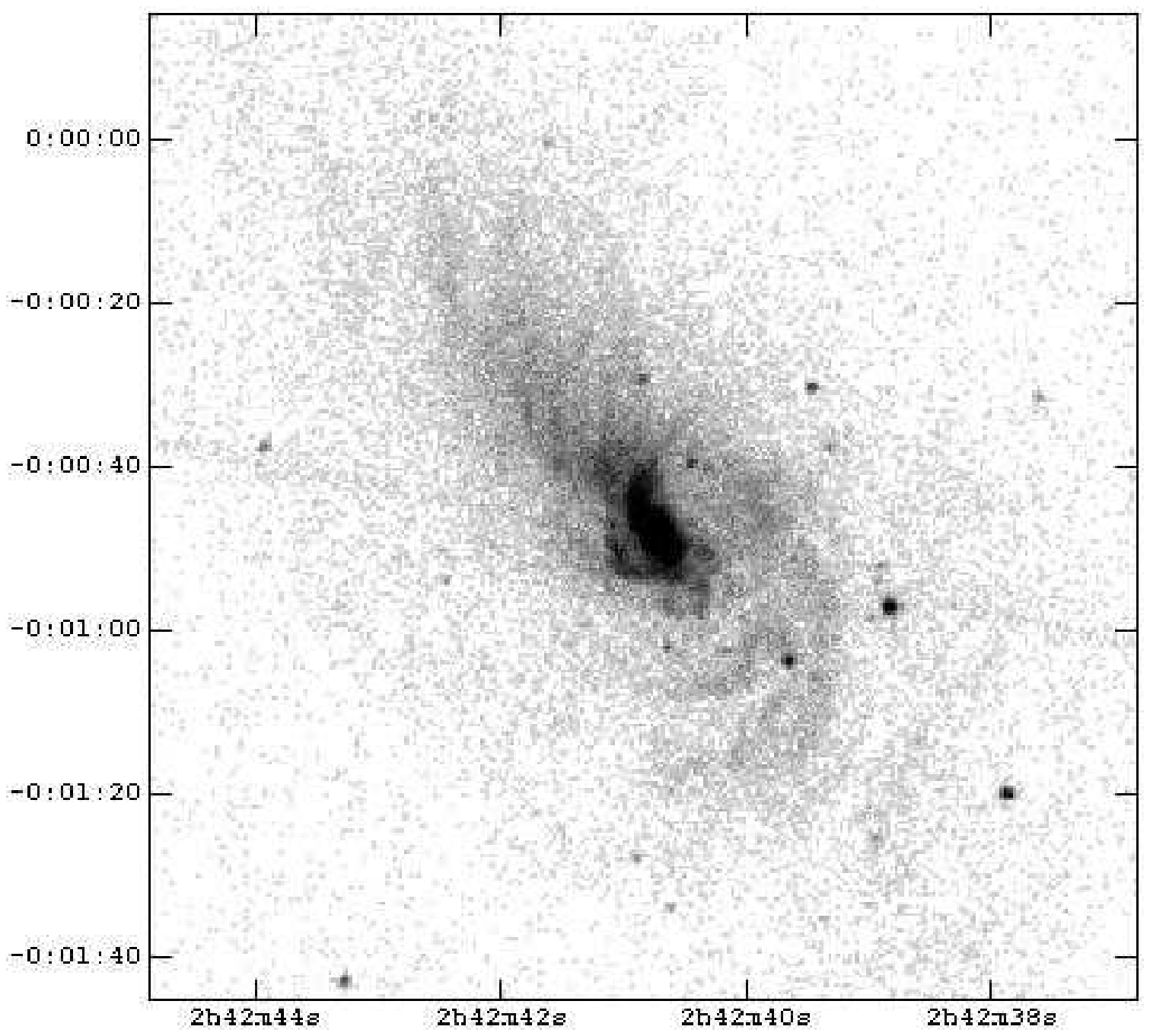,width=6.5cm,angle=0,clip=}
\hfill
\begin{minipage}[]{0.54\hsize}\vspace*{-6.5cm}    
\hfill
\psfig{file=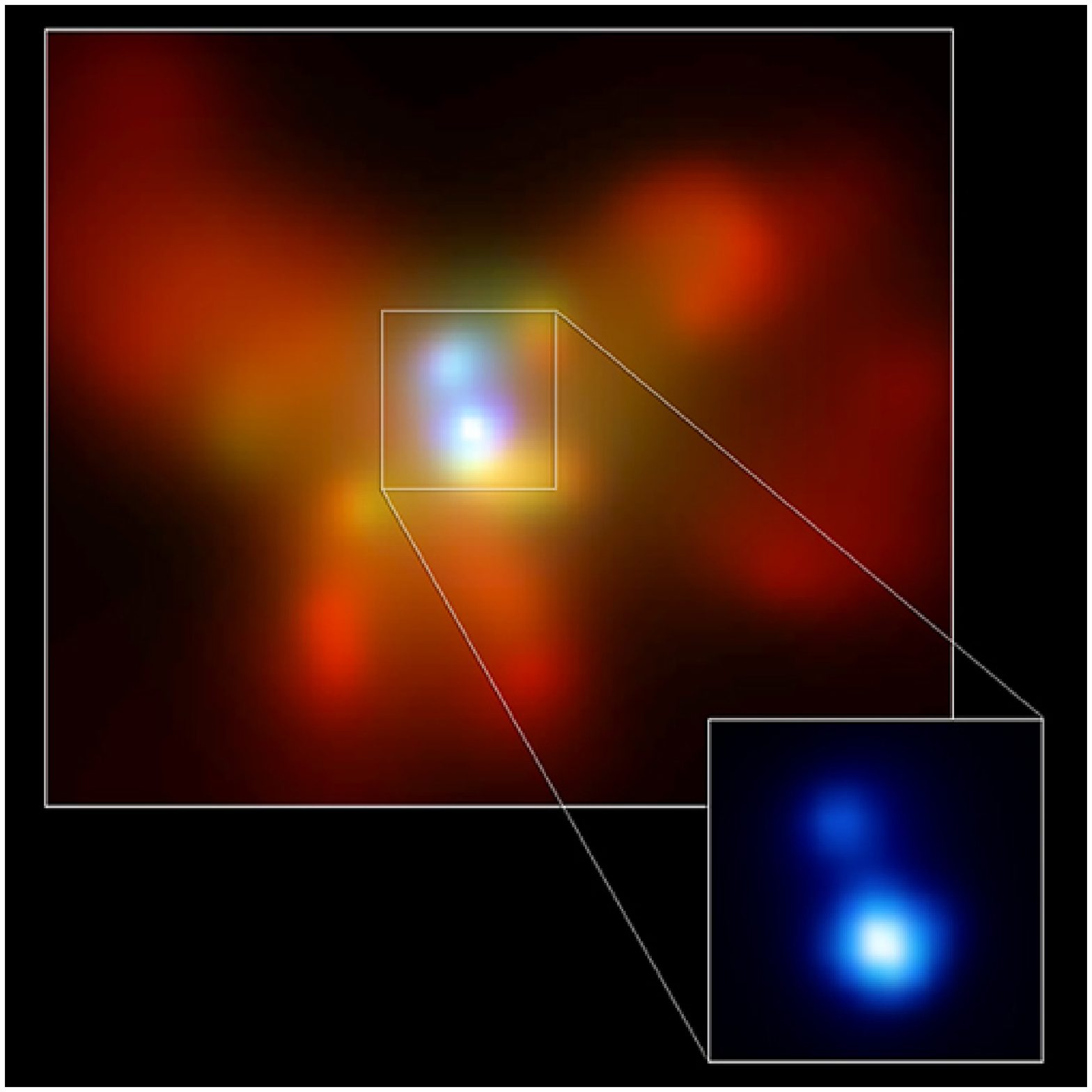,width=5.5cm,angle=0,clip=}
\end{minipage}

\caption[]{Left: {\sl Chandra} ACIS-S image of NGC\,1068
[taken from  Young, Wilson \& Shopbell, 2001].  Right:
{\sl Chandra} ACIS-S energy image of NGC\,6240 [Komossa et al. 2003];
the inset zooms onto the two hard nuclei.
 }
\vskip0.3cm

\psfig{file=komossa_f2a.eps,width=6.4cm,angle=-90,clip=}
\hfill
\begin{minipage}[]{0.64\hsize}\vspace*{-5.1cm}    
\hfill
\psfig{file=komossa_f2b.eps,width=6.4cm,angle=-90,clip=}
\end{minipage}
 \vspace*{-0.4cm}
\caption[]{X-ray spectrum of southern (left) 
and northern (right) nucleus of NGC 6240 
(the soft X-ray emission arises from
starburst-related extended emission, projected onto the nuclei).
In finer binning, a highly ionized iron line is revealed in addition
to the neutral line in both nuclei.  
}
\end{figure*}

\section{High-resolution X-ray spectroscopy of AGN}

For the first time, high-resolution spectroscopy of AGN  
can be performed in X-rays. With the exception of a few high-quality
{\sl ASCA} spectra of nearby AGN, the number of X-ray features
typically seen in AGN spectra was 2-3.
This strongly increased now; typically $\sim$20-30 features
are detected in the spectra of nearby AGN and even more in the brightest,
best-studied objects. 
A special role is played by warm absorbers.
Recognized earlier (see Komossa 1999 for a pre-{\sl Chandra} review), 
they have now become the most
important X-ray diagnostic of the central region of AGN,
imprinting many absorption lines on soft X-ray spectra of
Seyfert 1 galaxies. There have even been suggestions
that  the emission-lines detected in nearby Seyfert 2 galaxies
are also related to ionized absorbers, seen from the side.  

Basically, the trend emerged that Seyfert 1  X-ray spectra
are dominated by absorption, while Seyfert 2 X-ray spectra
are dominated by emission-lines. 
There are some exceptions, like NGC\,4151, a classical Seyfert\,1 galaxy
of type 1.5
(e.g., Osterbrock \& Koski 1976), 
which in X-rays looks more like a Seyfert\,2 (e.g., Kahn et al. 2001). 
Only very few AGN X-ray spectra turned out to be completely
featureless (e.g., Marshall et al. 2002).

\subsection{Seyfert 2 galaxies: emission-line spectroscopy}

I would like to start, again, with NGC\,1068. 
Both, {\sl Chandra} and
{\sl XMM-Newton} had a deep look
at this galaxy (e.g., Kahn et al. 2001, Kinkhabwala et al. 2002, Brinkman et al. 2002).
Its soft X-ray
spectrum is extremely rich in emission lines and is dominated by H-like and He-like
ions of low $Z$, and by Fe-L shell ions (Fig. 3).   
Kinkhabwala et al. (2002) inferred that the 
emission-line spectrum is photoionized by the nuclear continuum, and pointed 
out that the inferred column densities of the ions in the X-ray
gas match those derived for warm absorbers detected by absorption-spectroscopy
of Seyfert 1 galaxies.  
The relatively cold X-ray gas spatially coincides with the optical 
ionization cones of NGC 1068. The inferred broad
distribution of ionization parameters necessary to explain the X-ray
spectrum, requires the presence of a distribution of densities (several orders of magnitude)
at each radius in the ionization cone (Brinkman et al. 2002, Kinkhabwala et al. 2002).  
It is interesting to note,
that photoionization modelling of the {\em optical} radiation cone spectrum
of NGC 4151 (Schulz et al. 1993) and of the optical spectra of Seyfert 2 galaxies
(Komossa \& Schulz 1997) reached the same conclusion (i.e., the requirement of a 
range in densities at fixed radius in order to reproduce 
the observed emission lines).      

While most attention now is focussed on employing the new observatories 
in orbit, a number of surprises and exciting results may still linger in the archives
of {\sl ROSAT}, {\sl ASCA} and {\sl BeppoSAX} data. 
Making use of archival {\sl RXTE}, {\sl BeppoSAX} and {\sl ASCA}
data, Colbert et al. (2002) reported the detection of  a flare in the 6.7 keV
FeK emission-line of NGC 1068. The line is interpreted as arising in
warm reflecting material in the center ($<$ 0.2 pc from the AGN core)
of the galaxy. 

\begin{figure*}[h]
\psfig{file=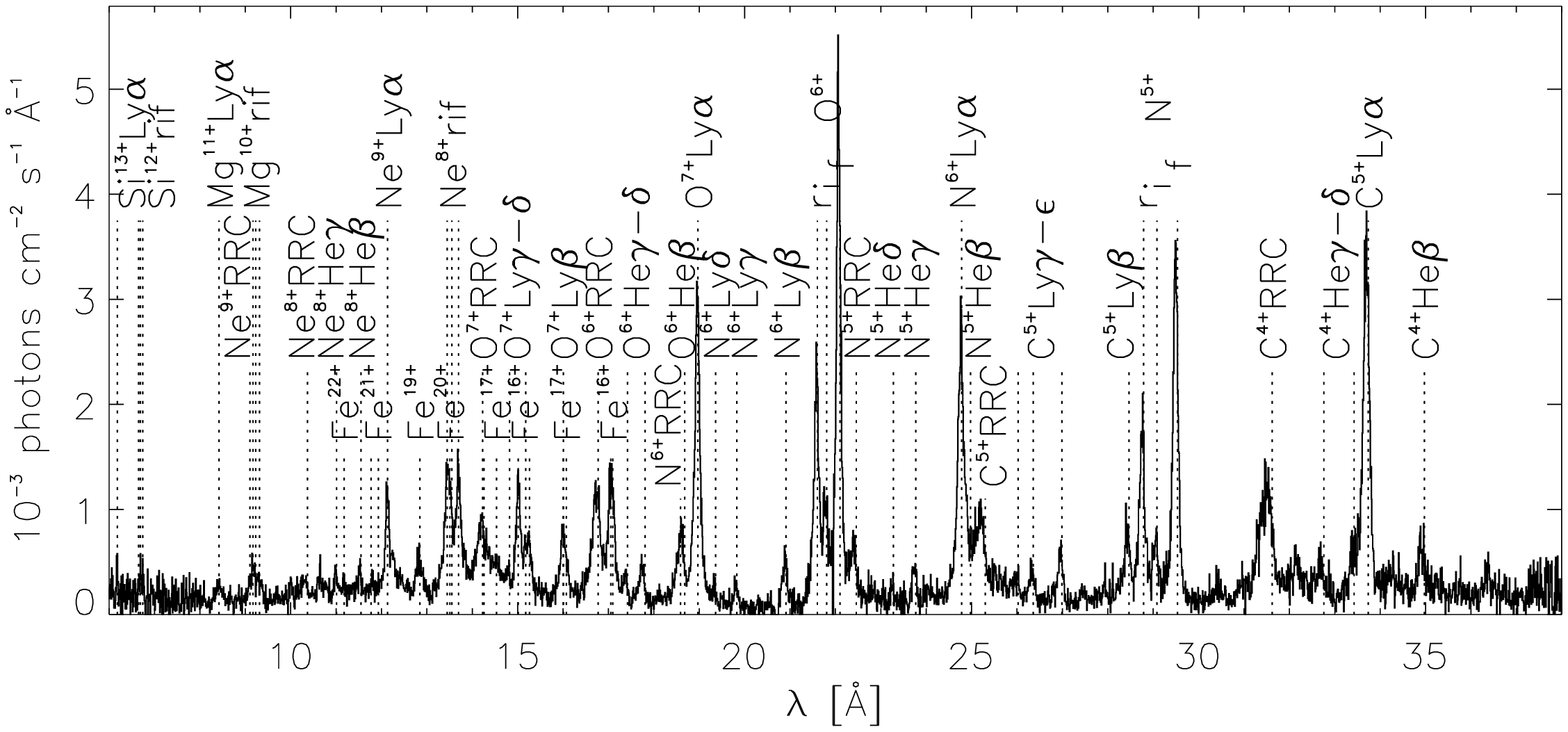,width=10.5cm,angle=90,clip=}
\psfig{file=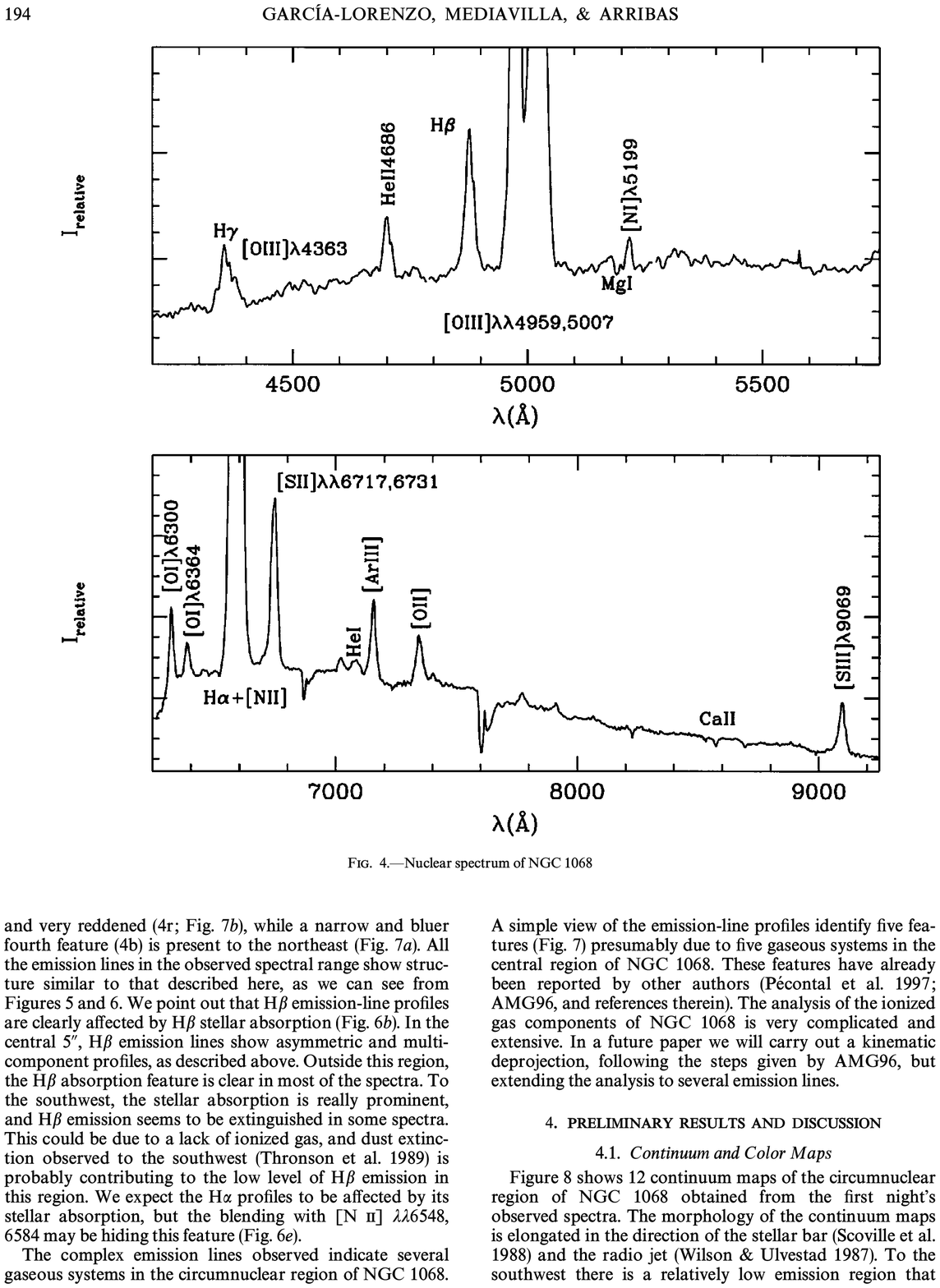,width=6.0cm,clip=}
\caption[]{ {\sl XMM} RGS spectrum of NGC\,1068
[taken from Kahn et al. 2001].  
The lower figure shows $\sim$50\% of a typical optical spectrum of this galaxy
[taken from Garcia-Lorenzo et al. 1999]. 
The comparison demonstrates that X-ray spectroscopy of nearby 
AGN is finally on its way 
to compete, or even beat, optical spectroscopy 
as far as the number of spectral features is concerned. }
\end{figure*}

The X-ray emission lines detected in the spectra of several more
Seyfert 2 galaxies (e.g., Mrk 3 and the Circinus galaxy)
and in NGC\,4151, 
contain important
information on the physical conditions in the line-emitting medium,
like temperature, density,
and the main gas excitation/ionization mechanism; photoionization
or collisional ionization.
Of particular importance in determining the
main power mechanism of the lines
are the Helium-like triplets, 
the widths of the radiative recombination continua, and the
strengths of the Fe-L complexes. 

The emission lines in the X-ray spectrum of NGC\,4151
(Ogle et al. 2000) are narrow and exhibit properties
characteristic for the narrow-line region. This is
the first X-ray NLR detected.{\footnote{Ultimately, the goal will
be to combine IR, optical, UV and X-ray observations in a renewed attempt
to finally derive the parameters the NLR of NGC 4151 and to infer
the unobserved EUV-SED of this galaxy; many attempts were done
during the last decade, demonstrating that several different solutions exist
(e.g., Contini et al. 2002 and ref. therein).}}

The general trend emerged, that the extended
gas in Seyfert\,2 galaxies appears
to be photoionized out to relatively large distances
from the nucleus.

\subsection{Seyfert 1 galaxies: absorption spectroscopy}

Neutral (`cold') or ionized (`warm') gaseous material is ubiquitous in
the AGN/SMBH environment, and therefore of utmost importance in
understanding the AGN phenomenon,
the evolution of active galaxies,
their link with starburst galaxies and ULIRGs, and the X-ray background.
X-ray absorption and emission features provide
valuable diagnostics of the physical conditions
in the X-ray gas. 

{\sl Chandra} spectroscopy of NGC\,5548, performed with the {\sl Low Energy Transmission
Grating Spectrometer}, LETGS, revealed for the first time a system
of deep narrow absorption lines in the X-ray spectrum of an AGN (Kaastra et al. 2000, 2002).
These lines arise from ionized material in the nucleus of NGC\,5548 and
show a range of ionization states.  

While all AGN in deep fields are exposed for megaseconds,
the record for the deepest grating observation of a nearby galaxy
is held by NGC\,3783.  The warm absorber
of this source was discovered in pioneering work by Turner et al. (1993).  
The 900\,ks {\sl Chandra} spectrum (e.g., Fig. 1 of  Kaspi et al. 2002) 
exhibits a wealth of absorption (and emission) features.

Using {\sl Chandra} and {\sl XMM-Newton},
absorption-complexes were detected and studied in several more 
AGN, including IRAS 13349+2438 (Sako et al. 2001),
NGC\,4051 (Collinge et al. 2001),
NGC\,3227 (Komossa et al. 2001), 
MCG\,$-$6-30-15 (Branduardi-Raymont et al. 2001, Lee et al. 2001),
Mrk\,509 (Yaqoob et al. 2002), NGC 4593 (Steenbrugge et al. 2002),
Ark 564 (Marshall 2002, Matsumoto et al. 2002) 
and NGC\,3516 (Netzer et al. 2002).
Generally, the trend emerges that the warm absorbers
are complex multi-component structures with a range in 
ionization parameters. 

The detailed atomic physics involved in producing the X-ray
spectra of Seyfert galaxies, and 
the consequences for our understanding of 
the central region of AGN, are still under scrutiny.
The X-ray spectra of the brightest AGN are so rich, that it will take
a while until all available information is extracted from them.  

\subsection{Iron lines}
The iron lines beyond 6 keV are the most important diagnostics
at these energies (see Fabian 2001, O'Brien et al. 2003 for reviews). 
If the line is formed in the inner parts of the accretion-disk,
its profile reflects 
the general relativistic effect of gravitational redshift
and 
the special relativistic effects of beaming and transverse
Doppler effect.
The best-studied iron-line case is MCG\,$-$6-30-15. 
At certain times, the red wing
of MCG$-$6-30-15 is very broad, extending down to very soft 
energies (Wilms et al. 2001).
  
{\sl XMM} and {\sl Chandra} confirmed and further corroborated
that the iron line profiles are complex
and that many components of the active nucleus may contribute 
to iron-line emission, including likely the BLR (NGC\,5548), the
torus (NGC\,3783, Mrk\,205), the X-ray ionization cone of NGC\,1068,
and a contribution from the outer parts of the accretion disk (MCG$-$6-30-15).
A particularly interesting case is the iron-line complex of NGC 3516.
Turner et al. (2002) reported the presence of several narrow components
within the FeK$\alpha$ profile of this galaxy. Their origin is still somewhat unclear. 
More details
on the iron-line topic are given in the contributions by J. Wilms  and T. Wang
(these proceedings).

\section{(Distant) quasars}

\subsection{BAL quasars} 

Broad absorption line (BAL) quasars are characterized 
by broad UV absorption lines.
It has been suggested that these lines arise in
a flow of gas which rises vertically from
a narrow range of radii from the accretion disk. The flow then bends
and forms a conical wind moving radially outwards (Elvis 2000).
Variants of radiatively-driven disk-winds were explored
(e.g., Murray et al. 1995, Proga et al. 2000, Proga 2001, Everett et al. 2002).
In some of these models, an X-ray absorber shields the wind downstream
from soft X-rays, allowing resonant-line driving to remain effective and
accelerate the outflowing BAL wind up to $\sim$0.1c.

Pre-{\sl Chandra/XMM} detections of BAL quasars in X-rays were rare.
Generally,
BAL quasars are X-ray weak, which is usually interpreted in terms
of strong excess absorption
(e.g., Green et al. 1995, Gallagher et al. 1999, Brinkmann et al. 1999,
Brandt et al. 2000, Wang et al. 2000).
{\sl Chandra} provided valuable new
constraints on the amount of absorption towards
selected BALs (e.g., Sabra \& Hamann 2001,
Oshima et al. 2001, Gallagher et al. 2002, Sabra et al. 2002), and an
{\sl XMM} observation let to a new identification of the X-ray counterpart
of the BAL quasar PHL 5200 (Brinkmann et al. 2002).
Almost all data still suffer from low S/N
(typically 50 to few hundred detected X-ray photons), though.
There are indications that the BAL material
is ionized instead of neutral.
This is definitely the case for the quasar APM 08279+5255
which has the
best-measured X-ray spectrum of any BAL quasar I am aware of.
APM 08279+5255 is magnified by a gravitational lense and
is among the most luminous objects in the universe (even after
correction for lensing).
A 100ks {\sl XMM-Newton} observation (Fig. 4) led to the detection
of a strong absorption feature of ionized iron, interpreted as K-edges,
arising from a warm
absorber of high column density (Hasinger et al. 2002).
A recent {\sl Chandra} spectrum
shows similar, but not identical features,  in the sense that the high-energy 
dip in the spectrum is interpreted/modeled in terms
of two highly blueshifted (in the quasar rest frame) iron-lines rather than edges
(Chartas et al. 2002). Combining the available observations,  
variability in the BAL-flow is implied. 

\begin{figure}[t]
\psfig{file=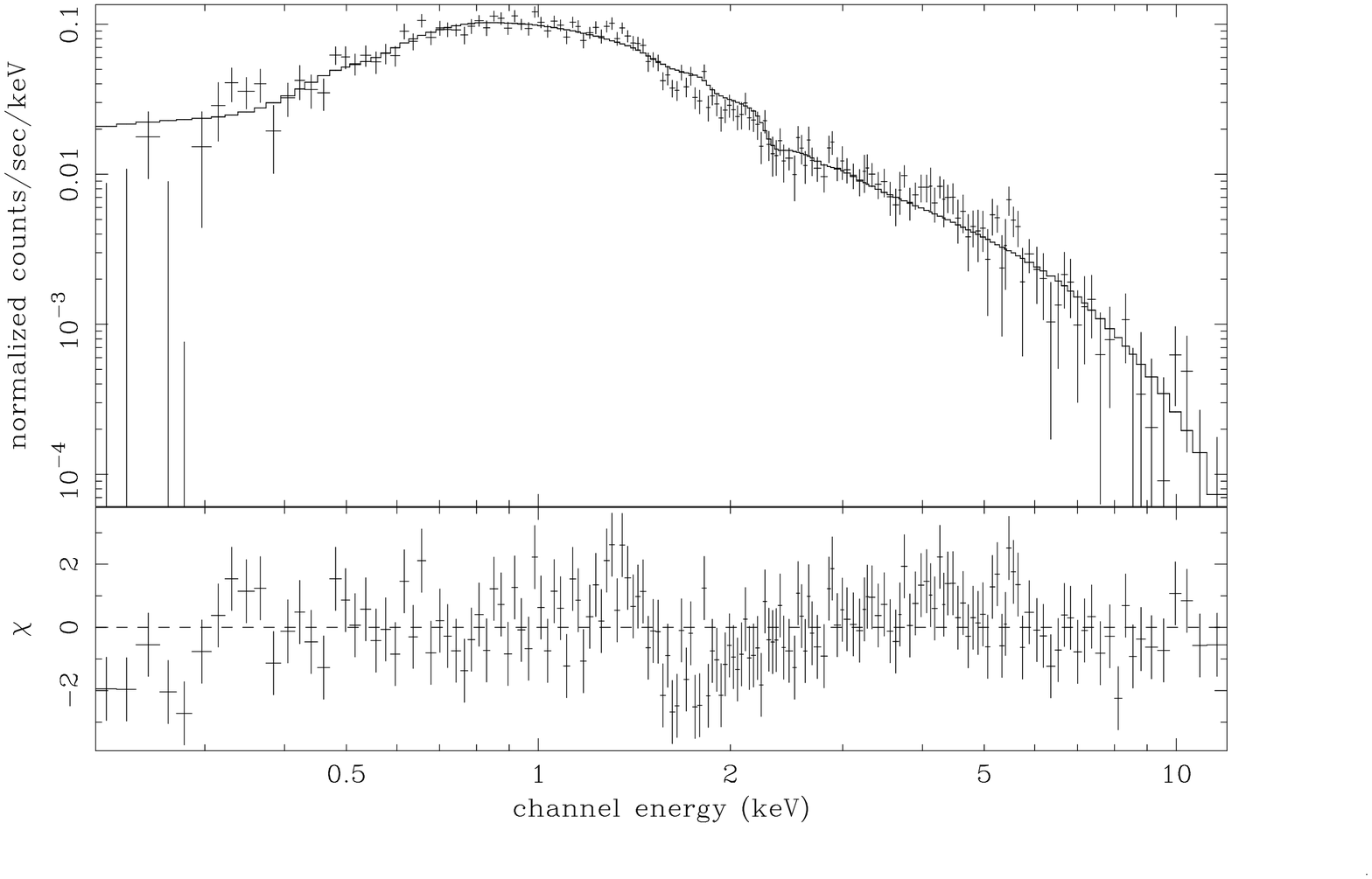,width=7.8cm,clip=}
\hfill
\begin{minipage}[]{0.49\hsize}\vspace*{-6.55cm}   
\hfill
\psfig{file=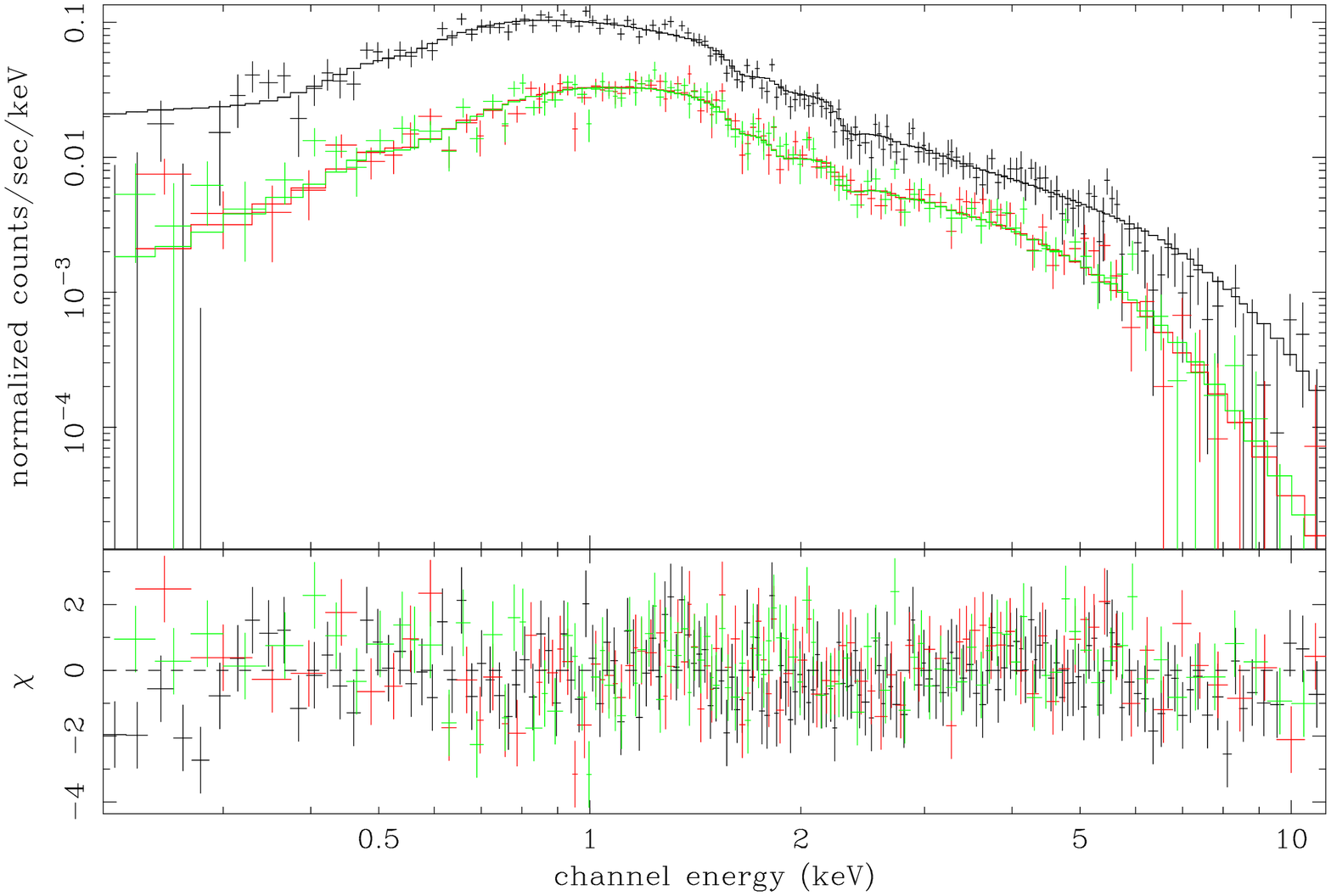,width=7.8cm,clip=}
\end{minipage}
\vspace*{-0.5cm}
\caption{ {\sl XMM-Newton} spectrum of the BAL quasar APM 08279+5255
at redshift $z$=3.91
(Hasinger et al. 2002). {\em Left:} {\sl XMM} EPIC-pn spectrum,
fit with a single powerlaw. An absorption feature
is visible at an energy corresponding
to ionized, redshifted iron.
{\em Right:} Combined EPIC and MOS spectra, fit with
a powerlaw plus an absorption edge of highly ionized iron.
}
\label{wa}
 \end{figure}

The inferred abundance of Fe/O $\simeq$ 3 $\times$ solar
is rather high, given the high redshift of
the object (Hasinger et al. 2002).
Taken at face value, it implies an efficient iron 
production mechanism in the early universe
and provides independent evidence for the presence of
a cosmological constant (see Fig. 4 of Komossa \& Hasinger 2003).

The unusual strength of the iron features and the indications
for strong variability make APM 08279+5255 an excellent target
for simultaneous, deep follow-up observations with {\sl Chandra}
and {\sl XMM-Newton}.

\subsection{Absorption in high-redshift quasars}

Evidence for excess X-ray absorption was found in
high-redshift, mostly radio-loud, quasars
(e.g., Wilkes et al. 1992, Elvis et al. 1994, Schartel et al. 1997, 
Yuan et al. 1998, Vignali et al. 2001).
The ionization state of the absorber remained largely unknown.
However, there is now growing evidence that these absorbers
are ionized, not neutral.
As shown by Schartel et al. (1997) the spectrum and spectral changes of
the high-redshift quasar PKS\,2351-154
($z$=2.67)
are well
explained by the presence of an ionized absorber
of column density $\log N_{\rm w} = 22.4$ which changes its ionization state
in response to intrinsic luminosity changes of the quasar.
PKS\,2351-154 is one of the very few high-$z$ quasars which show a {\em variable} UV
  absorption system as well.
For several years, this quasar
 held the record of being the most distant X-ray
warm-absorber candidate known,
recently exceeded by GB\,1428+42 and PMN\,J0525-33 (Fabian et al. 2001a,b).

Among the open questions related to absorption in (distant) quasars are:
What is the origin and nature of the high-z excess absorbers ?
Is this material preferentially warm or cold ? 
Why has it been more abundant in the past ? How does it evolve ?
  Why is excess absorption
mainly seen in high-redshift radio-{\em loud} quasars
whereas a number of (non-BAL) high-z radio-{\em quiet}
quasars appear to be absorption-free ?  Answers to these questions
are crucial for understanding the formation and evolution of AGN.
Apart from measuring ionic column densities, a very interesting
prospect is to determine element abundances in dust and gas
at high redshift.

\subsection{Quasars beyond redshift 5}

{\sl Chandra} and {\sl XMM} were used to search for and successfully
detect X-rays from
the highest redshift quasars,
identified in the course of the SDSS (e.g., Brandt et al. 2001, $z$=5.80; 
Mathur et al. 2002, Brandt et al. 2002, Bechtold et al. 2002, Schwartz 2002,
$z$=5.82, 5.99 and 6.28)
and discovered through the {\sl Chandra} multiwavelength project ChaMP (Silverman et al. 2002;
$z$=4.93), presently the most distant X-ray selected quasar. 
Most of the highest redshift quasars tend to be underluminous in X-rays.
It is presently being investigated whether this is due to evolutionary effects 
in the spectrum or the amount of absorption, or something else. The detailed analysis of
accreting black holes in the early universe,
i.e., the highest redshift quasars, will also be a major goal of future X-ray
missions.  

The status of the identification of 
many nearby and distant AGN 
in deep fields, and the implications, is summarized by G. Hasinger (these
proceedings) and Hasinger (2003). 

\section{Future X-ray missions}

While current X-ray missions like {\sl Chandra} and {\sl XMM-Newton} are
still expected to provide many new exciting results,
the new generation of X-ray survey missions will  
constitute a very useful supplement, by repeatedly re-scanning
the whole X-ray sky. This will lead to detections of numerous new X-ray
sources, and all kinds of unusual variability events among AGN.  
Among the planned missions are {\sl LOBSTER} (Fraser 2001), {\sl ROSITA}
(Predehl 2003), and {\sl MAXI} (Mihara 2001)
which will cover different energy bands with different spatial and spectral
resolution.   
In the long run, follow-ups of {\sl Chandra} and {\sl XMM} will come:
the Japanese missions {\sl ASTRO-E\,II} and {\sl NeXT} (Kunieda 2001),  
the US mission {\sl Constellation-X} (White \& Tananbaum 2001)
and the European mission {\sl XEUS} (Parmar 2003),
the latter two presently planned to be launched 
in the 2012-2015 time frame. These will combine
wide energy coverage with high sensitivity and spectral resolution
and with medium spatial resolution.

\vskip0.3cm
\noindent{\bf{Acknowledgments}} \\
It is a pleasure to thank the Max-Planck Gesellschaft  and the Chinese Academy of Science
for providing
travel funds in the course of the MPG-CAS exchange program
between China and Germany, and Wolfgang Voges for reading the manuscript. 
This and related papers are also available at: http://www.xray.mpe.mpg.de/$\sim$skomossa/

\end{document}